\documentclass[12pt,preprint]{aastex}

\newcommand{\mic}{\mbox{$\mu{\rm m}$}}

\newcommand{\gtappeq}{\raisebox{-0.6ex}{$\,\stackrel
{\raisebox{-.2ex}{$\textstyle >$}}{\sim}\,$}}
\newcommand{\kms}{\mbox{km s$^{-1}$}}

\shorttitle{Equatorial wind from S140 IRS 1}
\shortauthors{Hoare}

\begin{document}

\title{An equatorial wind from the massive young stellar object S140 IRS 1}

\author{Melvin G. Hoare}
\affil{School of Physics and Astronomy, University of Leeds, Leeds, LS2 9JT, UK} 
\email{mgh@ast.leeds.ac.uk}

\begin{abstract}
The discovery of the second equatorial ionized stellar wind from a
massive young stellar object is reported. High resolution radio
continuum maps of S140 IRS 1 reveal a highly elongated source that is
perpendicular to the larger scale bipolar molecular outflow.  This
picture is confirmed by location of a small scale monopolar near-IR
reflection nebula at the base of the blueshifted lobe. A second epoch
of observations over a five year baseline show little ordered outward
proper motion of clumps as would have been expected for a jet. A third
epoch, taken only 50 days after the second, did show significant
changes in the radio morphology.  These radio properties can all be
understood in the context of an equatorial wind driven by radiation
pressure from the central star and inner disc acting on the gas in the
surface layers of the disc as proposed by Drew et al. (1998). This
equatorial wind system is briefly compared with the one in S106IR, and
contrasted with other massive young stellar objects that drive ionized
jets.
\end{abstract}


\section{Introduction}

Mass-loss appears to be an integral part of the star formation
process. It is likely to play a major role in setting the final mass
of the star by reversing the infall and therefore in determination of
the IMF. Luminous ($>$10$^{4}$ L$_{\sun}$) young stellar objects
(YSOs) not only drive the well-known large-scale bipolar molecular
outflows (e.g. Lada 1985; Beuther et al. 2002), but also have small-scale
ionized stellar winds. These are weak ($S_{\nu}\la$ few mJy) and
compact ($\la$ 1\arcsec) radio sources (e.g. Tofani
et al. 1995). They are distinguished from ultra-compact H II regions
(UCHIIs) observationally by their radio spectral index that is usually
close to that expected for a stellar wind of +0.6 (Simon et al. 1983;
Wright \& Barlow 1975) compared to +2 for optically thick H II
regions. More conclusively, stellar wind sources have broad (few 100
kms$^{-1}$), optically thick, near-IR H I emission lines (Persson et
al. 1984; Bunn et al. 1995), whilst these lines in UCHIIs are
optically thin and a few 10s kms$^{-1}$ broad (e.g. Lumsden \& Hoare
1996). Theoretically of course, the origin of the stellar wind
material is from the star and/or disc itself, whilst the later UCHII
region phase begins when significant surrounding molecular cloud gas
has been ionized. Objects classified as hyper-compact H II regions
have intermediate properties and may represent a transition phase (see
review by Hoare et al. 2006).

The ability to resolve the radio emission from the stellar wind
sources
means that insights can be gained into the geometry of the
mass-loss. It is of particular interest to discover the relationship,
if any, between the ionized wind close to the star, and the much
larger scale bipolar molecular outflows that are invariably associated
with massive YSOs. In low-mass YSOs, the bipolar molecular outflows
are often accompanied by highly collimated stellar jets. These
optical/IR jets ends in bow shocks that appear to be responsible for
driving the molecular outflows in many cases (Chernin \& Masson 1995),
although not necessarily all (Lee et al. 2001). One may expect highly
collimated MHD driven jets to be less common in more massive stars
since magnetic fields are generally believed to play less of a role
compared to radiation pressure throughout their lives.

Indeed, evidence for such highly collimated jets from high-mass YSOs
has been much harder to come by (e.g. Poetzel et al. 1992). Searches
in the optical will obviously be hindered by the heavy extinction
close to the very embedded massive young stars, although searches for
shocked emission further out along the outflows have also shown no
signs of bow shock type emission (Alvarez \& Hoare 2005). Similarly,
near-IR observations of shocked molecular hydrogen do not often reveal
the jet and/or cavity structures seen in low-mass YSOs (Davis et
al. 1998), although see Davis et al. (2004). Radio investigations do
not suffer from extinction at all and the best example of a jet from a
massive YSO is the spectacular radio jet from GGD27, the exciting
source that drives the HH80-81 outflow. This parsec-scale precessing
radio jet which end in bow shocks was found by Mart\'{i} et
al. (1993). Follow-up multi-epoch studies by Mart\'{i} et al. (1998)
reveal proper motions of clumps in the jet, which correspond to
tangential velocities of at least 500 kms$^{-1}$. Such velocities are
similar to the FWZI of near-IR emission lines in other massive YSOs,
but unfortunately the exciting source of GGD27 cannot be seen directly
at near-IR wavelengths and no wind lines have yet been seen (Aspin
1994).

Another radio jet is that in Cep A2 (Rodr\'{i}guez et al. 1994) where
proper motion studies yield tangential velocities of 600 kms$^{-1}$
(Curiel et al. 2006).  It is perpendicular to a rotating disc seen in
sub-millimetre line and continuum emission (Patel et al. 2005) Some
aspects of its radio morphology do not fit the jet picture well (Hoare
\& Garrington 1995) and is not well aligned with the major CO outflow
axis in the region, although it is reasonably well aligned with an
outflow in HCO$^{+}$ (G\'{o}mez et al. 1999).  Other possible radio
jets from massive YSOs are GL 2591 (Campbell 1984; Trinidad et
al. 2003), W3 IRS5 d2 (Claussen et al. 1994; Wilson et al. 2003), W75N
VLA1 (Torrelles et al. 1997), and IRAS 20126+4104 (Hofner et al. 1999;
Trinidad et al. 2005). These radio jets are similar to the radio jets
that have been resolved in low-mass YSOs in several cases (Anglada
1996). It is then natural to suppose that the radio jets in massive
YSOs originate in more a scaled-up version of the MHD driven and
collimated flows invoked for low-mass objects.

However, by no means all massive YSOs possess radio jets. Observations
of S106IR, the exciting source of the famous bipolar H II region
S106, by Hoare et al. (1994) in fact showed the exact opposite
geometry. The elongation revealed by the MERLIN observations of this
source was aligned not with the outflow direction, but precisely
perpendicular to it, i.e. the radio morphology was disc-like rather
than jet-like. Of course, S106IR is somewhat of a unique object in
that it is already powering a well-developed H II region, which in
itself is unusual in having fast bipolar motions (Solf \& Carsenty
1982). Hence, there is a need to resolve the ionized winds of other
massive YSOs to see if such equatorial mass-loss is common or not. 

Another possible equatorial wind source is GL 490. Campbell et al.
(1986) resolved the 15 GHz emission from this source. It was clearly
elongated perpendicular to the CO outflow and was aligned with the
disc seen in later millimetre interferometric observations (Mundy \&
Adelman 1988; Schreyer et al. 2006), small scale reflection nebulosity
(Alvarez et al. 2004a) and spectropolarimetric data (Oudmaijer et
al. 2005). Peculiarly, this structure was not seen in deep 8 GHz VLA
observations by Gibb \& Hoare (in preparation). It is known that the
radio emission from massive YSOs is mildly variable in flux
(e.g. Hughes 1988) and so morphological changes have to be expected as
well.

Here radio maps are presented of the second confirmed incidence of
equatorial mass-loss from a massive YSO, namely S 140 IRS 1.  This
object is much more typical of the class of massive YSOs in that it
does not yet power an H II region like S106IR. It is the most luminous
of a cluster of objects seen at mid-IR (Beichman et al. 1979; Kraemer
et al. 2001) and near-IR wavelengths (Evans et al. 1989; Yao et
al. 1998). IRS 1 drives a strong bipolar CO outflow whose axis is in
the SE-NW direction (Hayashi et al. 1987; Minchin et al. 1993), the
cavity of which can be seen in the blueshifted SE lobe in C$^{18}$O
J=3-2 observations by Minchin et al. (1995). High
spatial resolution near-IR speckle observations reveal a monopolar
reflection nebula to the SE of the exciting source at the base of the
blueshifted outflow cavity (Hoare et al. 1996; Schertl et al. 2000;
Alvarez et al. 2004a).

The brightest of the IR sources in the embedded S140 cluster were also
found to be weak radio emitters by Beichman, Becklin \& Wynn-Williams
(1979) and further radio sources were seen by Evans et al. (1989).
High resolution radio observations were made by Schwartz (1989) which
revealed that IRS 1 was elongated in SW-NE direction for the first
time. At 5 GHz the core was about 1\arcsec\ long with weaker emission
extending a further couple of arcseconds to the SW. Schwartz (1989)
interpreted this morphology as a core-jet structure. He extended this
picture further by interpreting another diffuse patch of radio
emission 8\arcsec\ to the SW as a radio HH object. However, Schwartz
himself articulates the potential flaw in this picture since this
would mean that the jet is perpendicular to the outflow and thus
contradicting all known models of jet production. He raised the
possibility that the radio emission could originate in the ionized
surface of a disc. The observations presented here finally confirm
this disc-like morphology and interpretation. Tofani et al. (1995)
presented a high resolution 8 GHz VLA map of S140 IRS 1 which again
shows the elongated structure, this time with a significant curving of
the NE end around towards the N and possibly NW direction. The data in
this paper also recover this morphological detail.

\section{Observations}

S140 IRS 1 was first observed with a full 24 hour track with the full
MERLIN array of six telescopes on 7 and 8 June 1995. The continuum
frequency was 5 GHz with a bandwidth of 15 GHz in each
polarization. The maximum baseline was 218 km giving a potential
resolution of 50 mas. Frequent observations were made of the phase
calibrator 2221+625 (B1950 name) 0.7\degr\ away. The observations were
amplitude calibrated against OQ208 whose flux density was found to be
2.40~Jy in relation to 3C 286 whose flux density was assumed to be
7.38~Jy. Amplitude and phase corrections derived from mapping the
phase calibrator were interpolated and applied to S140 IRS 1. The
target was then mapped and CLEANed to produce the final images. In the
first epoch observations the phase centre was at $\alpha$=22$^{h}$
17$^{m}$ 41.5$^{s}$ and $\delta$=+63\degr 03\arcmin 47.5\arcsec
(B1950) to place it in the middle of the cluster of radio sources seen by
Schwartz (1989).

The second epoch observations took place on 3 March 2000 with exactly
the same setup and phase calibrator as before, but with the phase
centre now at S140 IRS 1 itself. The amplitude calibrator 0552+398
(B1950) was found to have a flux of 5.92 Jy.  These observations were
repeated seven weeks later on 21 and 22 April 2000. The amplitude
calibration of these observations appeared to be affected by rain
during the observation of 3C286 and so the flux of 0552+398 was
assumed to be the same as in the March 2000 observations. Maps were
made with a variety of reference antennae and time ranges to assess
the effect of the weather on the maps, but there was no significant
difference. The noise level in all the maps was about 60 $\mu$mJy per beam. 

\section{Results}

\subsection{First Epoch}

\begin{figure}
\plotone{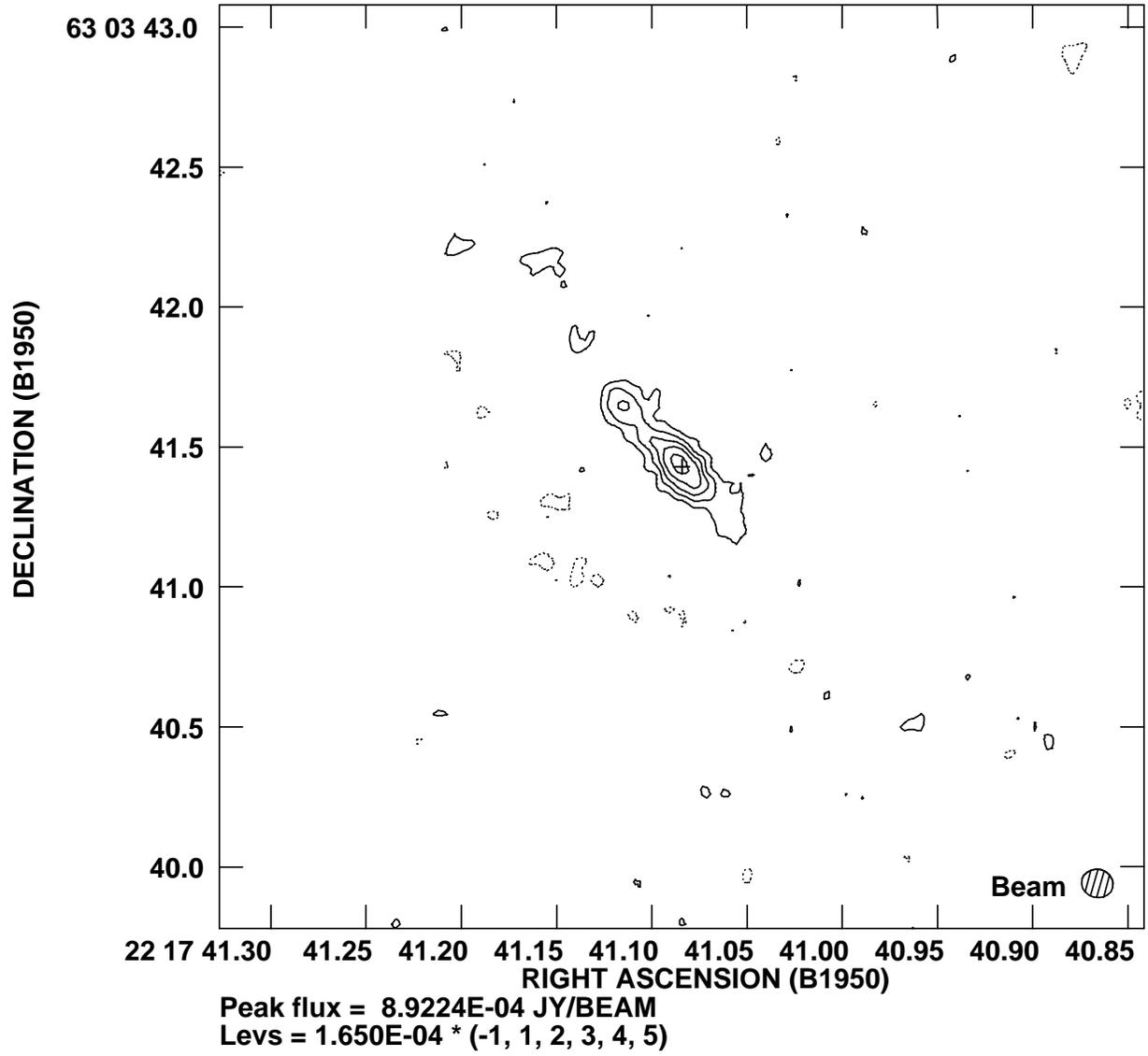}
\caption{First epoch 5 GHz MERLIN map of S140 IRS 1 with a
0\farcs114x0\farcs100 beam at PA=75\degr. This image was obtained on 7
June 1995. Lowest contour level is set at three times the noise level.}
\label{fig:ep1}
\end{figure}

The 5 GHz MERLIN map from the first observation in June 1995 is shown
in Figure \ref{fig:ep1}. A 5M$\lambda$ Gaussian taper has been applied
to the $uv$ data before making this map, the source being somewhat
over-resolved at the full 50 mas resolution. The noise level in the
map is 54 $\mu$Jy per beam. This is about 50 per cent higher than the
theoretically expected value, but that is to be expected since the
bright H II region S 140 itself is in the primary beam. The peak flux
of 0.89 mJy per beam corresponds to a peak brightness temperature of
3800~K which is similar to that found for the Herbig Be star MWC 297
(Drew et al. 1997), but lower than that for the other equatorial wind
system S106IR (Hoare et al. 1994). The integrated flux is
3.6$\pm$0.3~mJy.  A highly elongated structure is clearly seen at a
position angle (PA) of about 44\degr. It is barely resolved in the
minor axis and has a deconvolved axial ratio of about 7:1. The
parameters derived from a Gaussian fit to the source in the first
epoch are given in Table \ref{tab:params}.

\begin{deluxetable}{lcccrr}
\tabletypesize{\small}
\tablecaption{Parameters of S140 IRS 1 for the three epochs.\label{tab:params}}
\tablehead{
\colhead{Date}   & \colhead{RA} & \colhead{Dec} & \colhead{FWHM} & \colhead{PA} & \colhead{Flux}}
\startdata
 & 22$^{h}$17$^{m}$ & +63\degr03\arcmin & (\arcsec) & (\degr) &(mJy) \\ 
7/4/1995 & 41\fs0880$\pm$0\fs0014 & 41\farcs456$\pm$0\farcs010 &
 0.475$\pm$0.031$\times$0.124$\pm$0.008 & 43.7$\pm$1.5  & 3.6$\pm$0.3 \\ 
3/3/2000 & 41\fs0850$\pm$0\fs0021 & 41\farcs449$\pm$0\farcs015 &
 0.521$\pm$0.047$\times$0.149$\pm$0.014 & 43.1$\pm$2.3  & 3.4$\pm$0.5 \\ 
21/4/2000 & 41\fs0773$\pm$0\fs0009 & 41\farcs393$\pm$0\farcs007 &
 0.483$\pm$0.017$\times$0.124$\pm$0.004 & 43.7$\pm$1.5  & 3.6$\pm$0.3 \\ 
\enddata
\end{deluxetable}

The size and position angle agree well with the observations of
Schwartz (1989) and Tofani et al. (1995), but the structure is now
clearly seen at this high resolution. It is not a smooth structure but
appears to contain clumps, most notably the one that is 0\farcs30
NE of the peak. However, at this high interferometric resolution such
clumping can appear artificially enhanced. Further over-resolved
clumps appear to curve away to the NE and N; the most prominent being
at a position of 22$^{\rm h}$17$^{\rm m}$41\fs155 and +63\degr
03\arcmin 42\farcs18. These appear to follow the curving northerly
structure seen in the 8 GHz VLA map by Tofani et al. (1995) and it is
not surprising that this more diffuse structure is over-resolved in
the MERLIN map.

The total 5 GHz flux is significantly higher than the 2.2$\pm$0.4~mJy
quoted by Schwartz (1989) for the core observed in 1987. There is no
sign of the 'jet' that Schwartz saw to the SW with another
2.7$\pm$0.4~mJy, but it is quite a diffuse structure in the VLA map
and so is likely to be resolved out here. Tofani et al. found an
integrated flux of 7.9$\pm$0.7~mJy at 8.4 GHz. The spectral index of
the core from Schwartz (1989) is 0.94$\pm$0.23, which predicts a total
5 GHz flux of 4.8$\pm$0.6~mJy. This is somewhat higher than seen here,
but one would expect some of the flux seen by the VLA to be resolved
out. Hence, it is possible that there has been some mild flux
variability in S140 IRS 1 similar to that seen in other massive YSOs
like Cep A2 (Hughes 1988). However, it is difficult to draw firm
flux variability conclusions from interferometric observations taken
at different frequencies with different arrays.

\subsection{Multi-Epoch Observations}

\begin{figure}
\plotone{f2.eps}
\caption{Second epoch 5 GHz MERLIN map of S140 IRS 1 with a
0\farcs119x0\farcs104 beam at PA=-7\degr. This image was obtained on 3
March 2000. The cross marks the peak position in the first epoch image. 
Lowest contour level is set at three times the noise level.}
\label{fig:ep2}
\end{figure}

\begin{figure}
\plotone{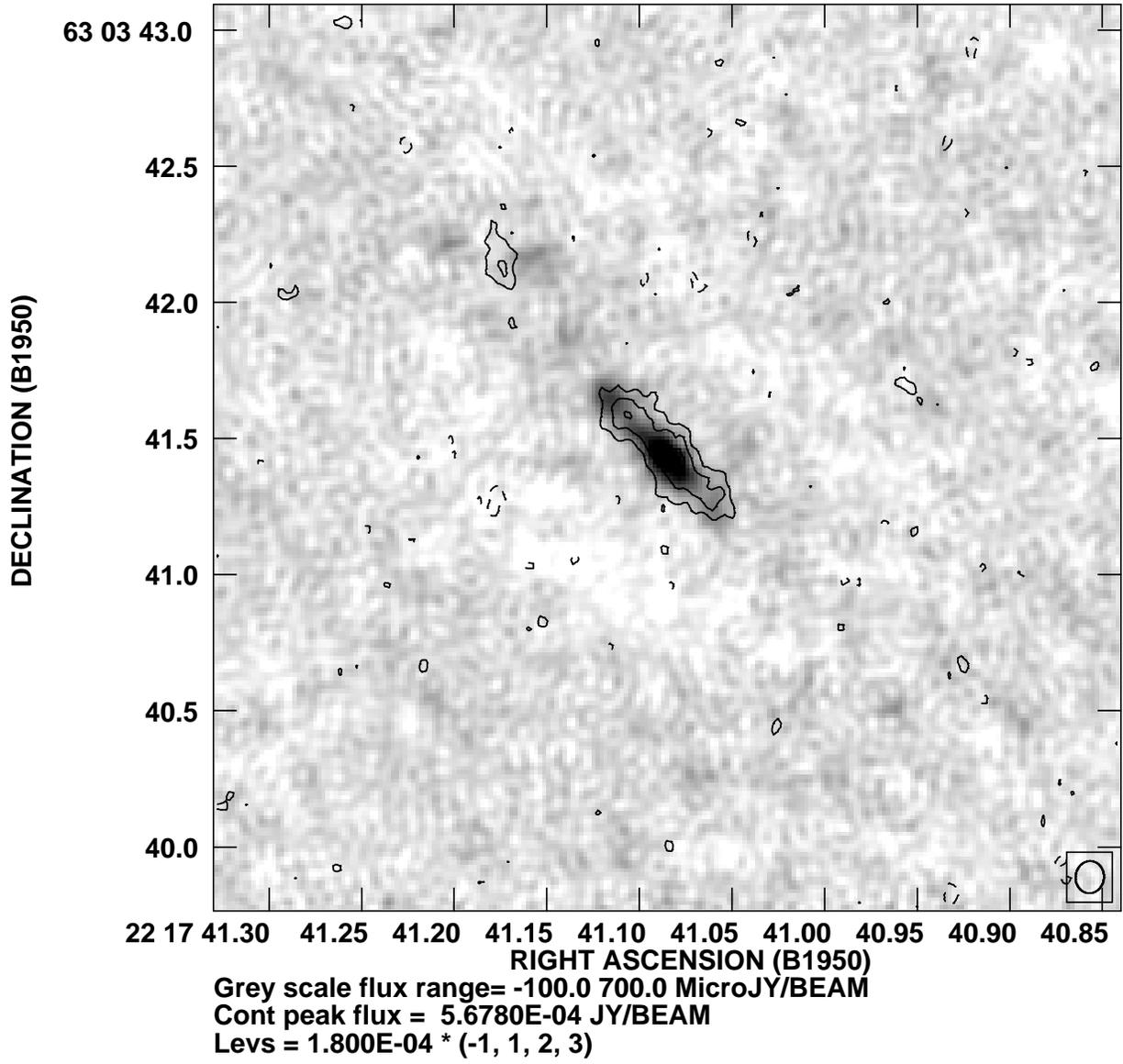}
\caption{Second epoch MERLIN map of S140 IRS 1 (contours) overlaid on
the first epoch MERLIN map (greyscale).}
\label{fig:ep12}
\end{figure}

To attempt to further clarify the nature of the wind emission in S140
IRS 1, second epoch observations were carried out over a 5 year
baseline sufficient to look for proper motions in gas that is moving
with velocities of order 100 kms$^{-1}$. The second epoch MERLIN image
is shown in Figure \ref{fig:ep2} and it is overlaid on a greyscale
version of the first epoch image from Figure \ref{fig:ep1} in Figure
\ref{fig:ep12}. A 4M$\lambda$ Gaussian taper has been applied when
making this image to deliver a similar size beam to that of the first
epoch. It can be seen from Figure \ref{fig:ep12} that the overall size
and structure of the source has remained the same and Table
\ref{tab:params} shows that the total flux has stayed
constant. However, there is now a much more even distribution of the
flux throughout the elongated core with the bright central core
present in the first epoch now less prominent. The overlay shows that
there has been no significant proper motion of the core with a
3$\sigma$ upper limit of 0\farcs049. At a distance of 800 pc this
corresponds to an upper limit in velocity of 39 \kms\ over the 5 year
baseline.

There has been significant movement of the patch of emission at
22$^{\rm h}$17$^{\rm m}$41\fs155, +63\degr 03\arcmin 42\farcs18, which
appears to have moved almost due east by 0\farcs15$\pm$0\farcs02.  If
this represents physical motion of a clump in the gas then this
corresponds to a tangential velocity of 120$\pm$16\kms. Even though
the measured velocity is comparable to the velocities inferred from
near-IR emission line profiles for massive YSO winds the direction of
motion is not that of a clump of gas moving radially outwards and is
not at all like the jet motions seen in other objects.  It is more
consistent with a straightening or opening out of the curved structure
seen in the first epoch and in Tofani et al.'s November 1992 VLA
image.  Since this likely to be the bright part of a larger,
over-resolved structure then it is also possible that the distribution
of material has remained the same but the illumination by ionizing
radiation has change between the epochs to highlight a different patch
(see below).

\begin{figure}
\plotone{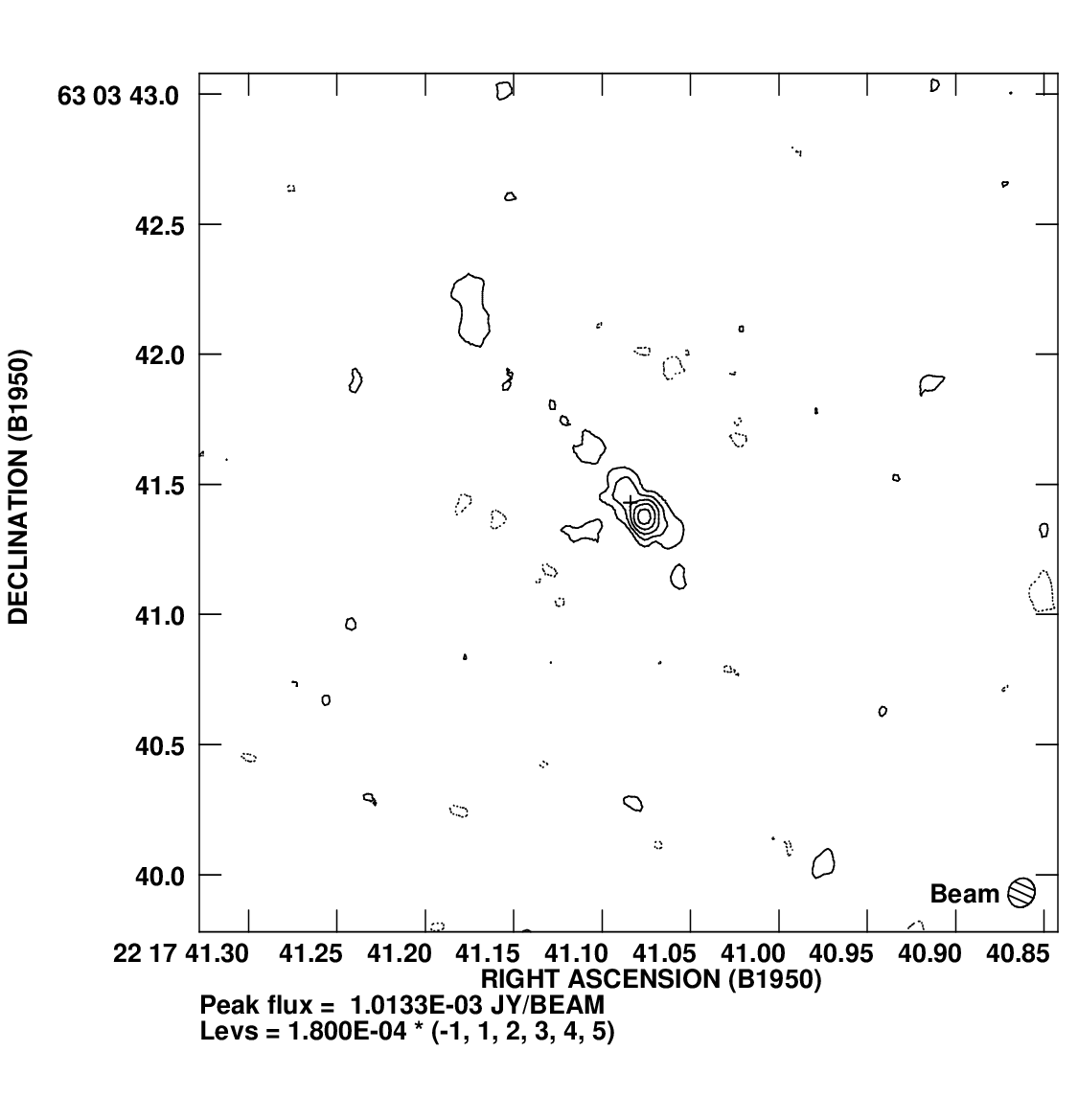}
\caption{Third epoch MERLIN map of S140 IRS 1 with a
0\farcs112x0\farcs100 beam at PA=-26\degr. This image was obtained on 21/22
April 2000. The cross marks the peak position in the first epoch image.
Lowest contour level is set at three times the noise level.}
\label{fig:ep3}
\end{figure}

Figure \ref{fig:ep3} shows the MERLIN map just 50 days after that in
Figure \ref{fig:ep2}. Here the changes in morphology of the source are
much more dramatic. Several checks were made to see if the periods of
rain during the observations were affecting the resulting source
morphology, but they did not. Small changes in morphology can occur
due to different {\it uv} coverage of the same source, but all the
observations were very similar full tracks.
In the third epoch the emission is concentrated in a
compact region to the SW of the original core position, although again
the integrated flux has stayed remarkably constant (Table
\ref{tab:params}). Emission in the NE part of the elongated core is
now almost absent. The patch of emission which showed significant
proper motion over the previous 5 years is unchanged over this much
shorter interval. This is as expected and bolsters the significance
of the other changes in morphology.

\section{Discussion}

\begin{figure}
\plotone{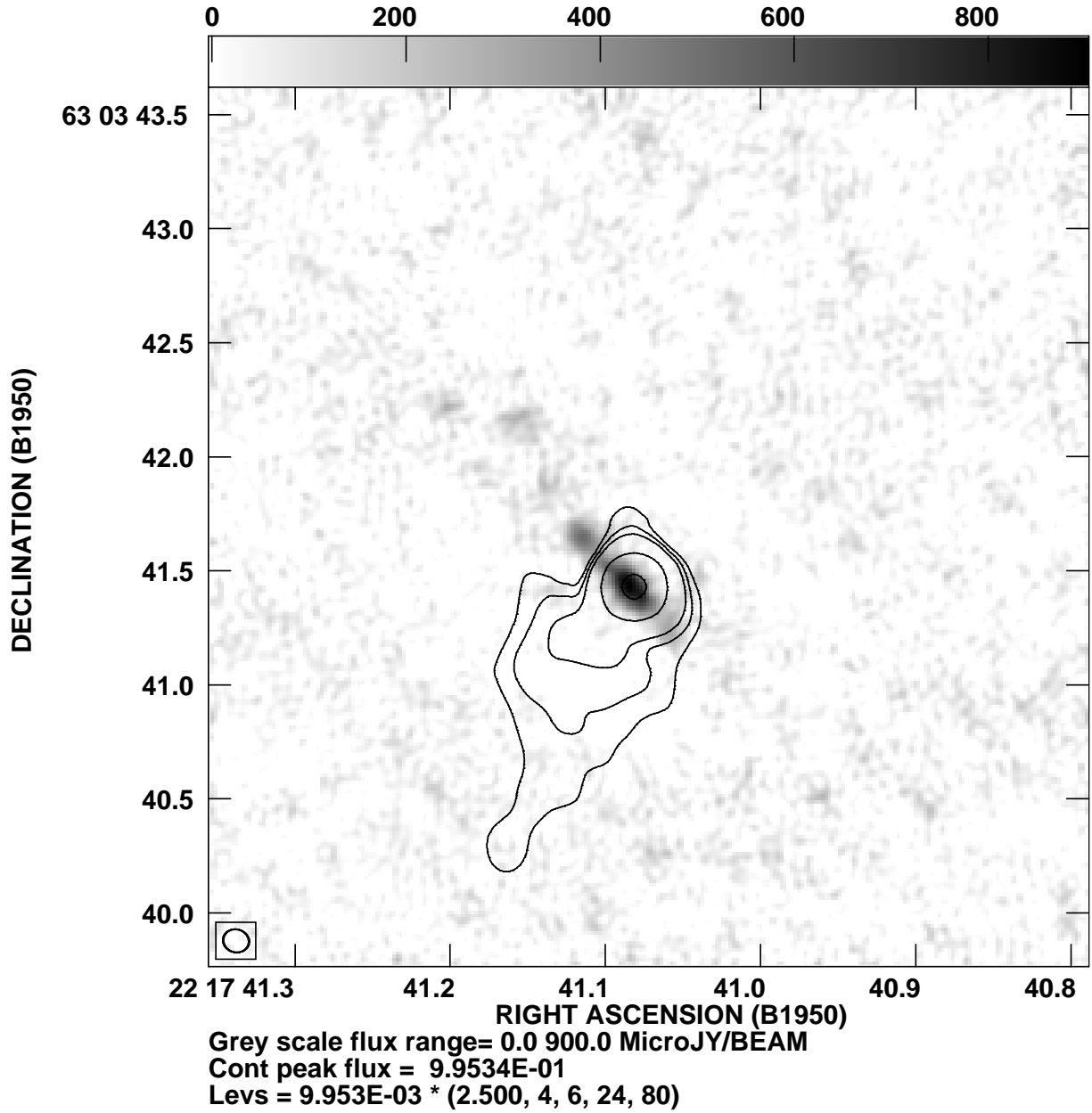}
\caption{MERLIN 5 GHz map from Figure 1 (greyscale) overlaid with
near-IR K-band reconstructed speckle image with a resolution of
0\farcs2 (contours) from Hoare et al. (1996).}
\label{fig:5K}
\end{figure}

So, what is the nature of the radio emission from S140 IRS 1? The
highly elongated, clumpy and curving structure is highly reminiscent
of the jet from the luminous YSO GGD27. However, as noted by Schwartz
who first put forward the jet interpretation, there is a major problem
with this in that such a jet would be perpendicular to the large scale
bipolar molecular outflow in the region, which appears to be powered
by IRS 1. The true nature of the radio source is revealed through high
resolution observations in the near-IR. Figure \ref{fig:5K} shows a
greyscale of the MERLIN map from Figure \ref{fig:ep1} overlaid with
contours of near-IR K-band (2.2\mic) emission from a reconstructed
speckle image obtained by Alvarez et
al. (2004a). The unresolved near-IR point source as been arbitrarily
aligned with the peak of the radio emission given the lack of accurate
near-IR astrometry. The position of the 2MASS counterpart of IRS 1.
is located 0.2\arcsec\ E and 0.8\arcsec\ S of the peak MERLIN position.
Given that the extended nebular emission dominates over the point source
in the K-band (Alvarez et al. 2004a) this is consistent with the
adopted alignment within the errors of 2MASS, the different resolutions,
techniques and optical to radio frame correspondence.

The diffuse near-IR light forms a monopolar nebula to the SE of the
star, which is interpreted as a reflection nebula arising in the
outflow lobe that is tilted towards us. Note that the blueshifted CO
outflow lobe is the one to the SE. The scattered nature of this nebula
has been confirmed by a deeper and higher resolution speckle
polarimetric image taken by Schertl et al. (2000). This reflection
nebula aligns well with the larger scale molecular outflow as is
confirmed by a high resolution interferometric CO map obtained by
Gibb, Hoare \& Shepherd (in preparation), which clearly
shows IRS 1 to be at the centre.

It is now clear that the radio emission from S140 IRS 1 arises in a
disc-like structure perpendicular to the bipolar outflow rather than
in a jet. Hence, it becomes the second definite case of an equatorial
ionized wind system to be discovered in a luminous YSO after S106IR.
A promising model for such an equatorial wind has been developed by
Drew et al.(1998). Here the radiation pressure from the star
and accretion disc can lift and accelerate material from the surface
of the disc and blow it away in a predominately lateral direction. The
velocities obtained for the gas where the majority of the emission is
expected are of the order of a few hundred kms$^{-1}$ in agreement
with the widths of the near-IR emission lines in massive YSOs. 

The large changes in morphology, but not flux over the short time
interval between the second and third epoch MERLIN observations cannot
be explained by any motion of clumps in the wind since the implied
velocities would be much too fast. These rapid changes are consistent
with a picture in which small changes in illumination by the ionizing
source produce changes in the winds structure. Such shadowing effects
are greatly enhanced in the foreshortened site lines of a central star
viewed from the surface of an accretion disc. Furthermore, it is known
that radiation-driven winds are inherently unstable (Owocki et
al. 1988), which can produce sudden large changes in the illumination
and driving force. Hence, such rapid changes in structure could be
expected from a radiation-driven disc-fed wind model.

It is unlikely that the radio emission arises simply from the
photoionized surface of the disc. The kinematics would then be
dominated by disc rotation, but the near-IR emission line profiles do
not appear to be easily explained by a simple accretion disc
explanation, e.g. there is no sign of double-peaked profiles (Drew et
al. 1993; Bunn et al. 1995).  Even the first attempt to model the
near-IR line profiles from the radiatively driven disc wind model of
Drew et al. (1998) predicts double-peaked rather than the observed
single-peaked profiles implying too high a rotational component of the
motion relative to radial motion (Sim et al. 2005).  In any thin disc
explanation the extreme axial ratio of MERLIN observations would imply
a system very close to edge-on, i.e. $i\gtappeq 82\degr$. At face
value this would be at odds with the monopolar reflection nebula seen
in the near-IR since this requires an intermediate inclination angle
so that light scattered from the redshifted lobe is blocked by the
inclined disc or torus (Alvarez et al. 2004b).  However, the sparse
nature of the MERLIN array may well resolve out diffuse emission on
the minor axis. The much more fully sampled VLA map of Tofani et
al. (1995) has an axial ratio of about 2:1 and hence is more consistent with
the near-IR picture.

In the context of the equatorial wind scenario, the extended 'jet'
feature to the SW of the main S140 IRS 1 source seen by Schwartz
(1989) needs to be re-assessed. Its nature is unclear in this
scenario, but it is does appear to get much more significant at lower
frequencies. The much flatter spectrum implies optically thin emission
or even a shocked origin. Perhaps the equatorial wind or UV radiation
field is illuminating an extended disc or torus, but it is not clear
why this would occur only on one side of the object. A similarly
perplexing unexplained feature is present at low frequencies in the
massive YSO Cep A2 (Hoare \& Garrington 1995). As for the radio source
that Schwartz interprets as a 'bullet' some 7\arcsec\ to the SW of IRS
1, its position angle from IRS 1 is actually 37\degr\ and hence some
way off being aligned with the elongated radio source. It also
coincides with what appear to be shocked structures in the speckle
image of Weigelt et al. (2002) that are not necessarily assoicated
with IRS 1 at all.

A high resolution 43 GHz VLA continuum map of S140 IRS 1 shows the
same elongated morphology, but at a greater position angle (Hoare
2002; Gibb \& Hoare 2006 in preparation). It also has a slight twist
giving it a point symmetric appearance which is often interpreted as
rotation or precession. The emission from these higher frequencies
will originate from deeper within the wind closer to the star.  The
sense of implied rotation, being counter-clockwise, is also consistent
with a similar origin for the feature that curves off to the north in
Fig \ref{fig:ep1} and Tofani et al.'s 8 GHz image. More speculatively
it can be seen that the brighter part of the near-IR reflection nebula
is on the southern side. This would be consistent with an outflow
cavity illuminated by a disc inclined further in the counter-clockwise
direction like in the well-studied Chameleon Infrared Nebula (Gledhill
et al. 1996). There is evidence of such rotational motions in other
massive YSOs. A possible change in position angle over time is seen
for the jet in Cep A2 (Curiel et al. 2006).  For this object it is
interesting to note that the larger scale pattern of optical and
near-IR bow shocks in the Cep A East outflow are also consistent with
a clockwise rotation of the jet over time. They show a sequence of
decreasing length (age) at PAs of 77\degr, 66\degr\ and 54\degr\
(Corcoran et al. 1993; Hiriart et al. 2004), although this must have
occurred over much longer timescales.

One possible counter to the equatorial wind picture painted for S140
IRS 1 above is the discovery by Weigelt et al. (2002) of three bow
shock structures seen to the NNE of IRS 1 that also appear to be
driven by IRS 1 or a source close to it. They discuss various
possibilities to explain this and conclude that a binary scenario with
two mis-aligned flows is most likely. Preibisch \& Smith (2002) take
this further claiming that the radio source in IRS 1 is the driving
source for these structures and therefore returning to a jet
interpretation. The inferred rotation of the jet driving the three
structures is the opposite to that inferred here for the radio source.
Since the bow shocks are much weaker than the outflow indicators in
the SE-NW direction they are most likely driven by
a less luminous source in the vicinity of IRS 1. Thus the association
of the radio source with the dominate luminosity source and outflow
reasserts the disc-like interpretation.

\section{Conclusions}

Multi-epoch high resolution radio continuum mapping of S140 IRS 1 have
shown it to be the second known example of an equatorial ionized wind
from a massive YSO. The highly elongated 5 GHz emission is
perpendicular to both the large scale bipolar molecular flow and the
small scale monopolar near-IR reflection nebula at the base of the
blueshifted lobe. Proper motion studies reveal no jet-like motions of
clumps in the wind. Instead there is an apparent flattening of the
curved structure on the NE side over 5 year baselines. On 50 day
timescales there are also significant changes in the distribution of
the radio emission whilst the total flux remains constant.  The radio
properties of S140 IRS 1 can be understood in terms of of the
radiatively driven disc wind model put forward by Drew et al. (1998).
There are some indications that the position angle of the disc is
changing with time.

The question of why this source and S106IR have equatorial ionized
winds whilst other massive YSOs such as GGD 27, Cep A2, G35.2N and
IRAS 16547-4247 have ionized jets is not known. These objects all have
many observational characteristics in common, including their
luminosity. One thing that does set the jet sources apart is that they
are more embedded and are not seen directly in the near-IR at all.
Indeed, the central source of IRAS 16547-4247 is not even visible in
the Spitzer GLIMPSE survey images.  This could indicate an
evolutionary trend in the younger, more embedded sources have jets,
and the equatorial disc wind sets in later as the envelope is
dispersed. In theoretical terms this could be interpreted as the
magnetic fields required to produce highly collimated jets decaying
with time and radiation pressure taking over. However, the
line-of-sight extinction and near- to mid-IR colours are highly
viewing angle dependent (Yorke \& Bodenheimer 1999; Yorke \&
Sonnhalter 2002; Whitney et al. 2003) and cannot be taken as a good
indication of age.  Alternatively, the wind morphology and mass-loss
mechanism could be set by the initial conditions such as angular
momentum and magnetic fields.

To distinguish between the different scenarios to explain the
dichotomy in wind morphology will require a much larger sample of
massive YSOs to be resolved in their radio continuum. This is a task
for the EVLA and e-MERLIN arrays. It would also be advantageous to
acquire complementary velocity resolved IR H I line profiles. These
yield information on the acceleration of the gas close to the source.
Unfortunately, most of the jet sources are too deeply embedded to use
the traditional Brackett series lines and longer wavelength probes
will likely have to be deployed. Combined with multi-dimensional
modelling of both the line and continuum data these data should then
allow the physics behind the ionized mass-loss in massive YSOs to be
unlocked.

\acknowledgements

MERLIN is a National Facility operated
by the University of Manchester on behalf of PPARC.  The
staff at Jodrell Bank Observatory are thanked for their assistance in making the
observations with MERLIN. In particular, Drs Simon Garrington and Tom
Muxlow provided invaluable help with the data reduction.  Useful
discussions on the interpretation of the data were held with Dr Janet
Drew. The referee is acknowledged for improving the clarity of the paper.

{\it Facilities:} \facility{MERLIN}

\end{document}